\documentclass[epsfig,12pt]{article}
\usepackage{graphicx,epsfig}
\usepackage{amsmath,amssymb}
\def\to{\rightarrow}

\begin{document}
\begin{titlepage}

\begin{center}
\Large{{\bf Unitarity Constraints on effective interaction in $\pi
N$ scattering}} \vskip 1cm

\normalsize {Mingxing Luo\footnote{Email: luo@zimp.zju.edu.cn}, Yong Wang\footnote{Email: porland@zju.edu.cn}
and Guohuai Zhu\footnote{Email: zhugh@zju.edu.cn} \\
\vskip .5cm
 {Zhejiang Institute of Modern Physics, Department of Physics, \\
 Zhejiang University, Hangzhou, Zhejiang 310027, P.R. China}
  \\
 }
\begin{abstract}
Positivity constraints are derived on pion-nucleon scattering amplitudes and
their even-order derivatives inside the Mandelstam Triangle with the help of dispersion relations.
Fairly interesting constraints are obtained on some of the low energy constants,
by a combination of the chiral perturbation theory for heavy baryons
and existing fitting results from available pion-nucleon phase shifts at intermediate energies.
\end{abstract}
\vskip 1cm
\end{center}

\end{titlepage}
From the viewpoint of effective field theory(EFT),
new physics effects can be parameterized as Wilson coefficients(WCs) of higher dimensional effective operators.
Usually, these WCs are assumed to be totally free.
Of course, concerning the suppression of high energy scale,
it is generally believed that these WCs cannot be arbitrarily large,
which is also required by the applicability of perturbation methods, for the theory to be of practical use.
Were the fundamental theory known, these coefficients should be, at least in principle, uniquely determined
as the low energy limit of the underlying one.
Naturally one may ask, as there is currently no clear clue about the underlying theory,
whether it is still possible to give some constraints -though maybe rather weak in many cases-
on the WCs of higher dimensional effective operators.

Along this line of thought, Adams {\it et al.} \cite{Adams:2006sv}
recently investigated positivity constraints on the WCs of a set of
effective operators by assuming causality, analyticity and unitarity
of the ultraviolet theory. In chiral perturbation theory, the same
assumptions lead to well-known dispersion relations. From the
viewpoint of EFT, Distler {\it et al.} \cite{Distler:2006if}
reconsidered implications of dispersion relations as imposing bounds
on coefficients of higher dimensional effective operators in vector
boson scatterings. In this paper, we will study from this angle
implications of dispersion relations on effective pion-nucleon
interactions. Here the underlying theory is supposed to be known
(quantum chromodynamics) but difficult to be applied directly. So
our results will be of help to the analysis of low energy hadronic
processes, in addition to illustrating the general principle of this
line reasoning. Combining with fitting results \cite{Fettes:1998ud}
from available pion-nucleon phase shifts at intermediate energies,
potentially strong constraints are obtained on some of the WCs of
higher dimensional operators.

Low energy pion-nucleon scatterings can be described by the following effective Lagrangian \cite{Fettes:2000gb}
\begin{eqnarray}
{\cal L}_{\pi N}=\bar{\Psi} \left ( i {\rlap{\hspace{0.06cm}/} {D}}
- m + \frac{g_A}{2}{\rlap{\hspace{0.02cm}/} {u}} \gamma_5 \right )
\Psi + \sum \limits_{i=1}^{7} c_i \bar{\Psi} O_i^{(2)} \Psi + \sum
\limits_{j=1}^{23} d_j \bar{\Psi} O_j^{(3)} \Psi + \sum
\limits_{k=1}^{118} e_k \bar{\Psi} O_k^{(4)} \Psi + ...~,
\end{eqnarray}
where $O_i^{(l)}$s denote dimension-$l$ chiral operators, and their
corresponding Wilson coefficients $c_i$, $d_j$, $e_k$ are often
called low energy constants (LECs), with units of $\mbox{GeV}^{-1}$,
$\mbox{GeV}^{-2}$ and $\mbox{GeV}^{-3}$, respectively. The ellipsis
denotes operators with chiral dimension five or higher. The readers
may refer to Ref. \cite{Fettes:2000gb} and references therein for
explicit expressions of $O_i^{(l)}$.

Dispersion relations for $\pi N$ scatterings have been known for several decades \cite{Hohler83, Hamilton63}.
For our purpose, we will only consider dispersion integrals with $t=0$.
At this limit, the imaginary part of a scattering amplitude is proportional to
the corresponding cross section through the optical theorem.
This may provide the required property of positivity which in turn gives constraints on the LECs.
Inside the Mandelstam triangle,
it is straightforward to rewrite fixed-$t$ dispersion relations \cite{Buettiker:1999ap}
in terms of scattering amplitudes $T^I (s, t)$:
\begin{eqnarray}\label{dispersion}
\frac{\partial^n T^{I}(s,0)}{\partial s^n}=\frac{n !}{\pi}\int
\limits_{(m+M)^2}^{\infty} dx \left [ \frac{\mbox{Im}
T^{I}(x,0)}{(x-s)^{n+1}} - \frac{\mbox{Im} T_u^{I}(x,0)}{(u-x)^{n+1}}\right
] + \frac{\partial^n P^{I}(s,0)}{\partial s^n}~,
\end{eqnarray}
with
\begin{equation}
P^{3/2}=\frac{g^2}{m}\frac{(s-m^2-M^2)}{(u-m^2)}~,\hspace*{0.5cm}
P^{1/2}=\frac{g^2}{2m}\frac{(s-m^2-M^2)(4m^2-3u-s)}{(m^2-u)(m^2-s)}~.
\end{equation}
Here the superscript $I$ denotes the total isospin of the pion-nucleon system
and we have exploited the fact that the scattering amplitudes are real inside the Mandelstam triangle.
Parameters $m$ and $M$ represent masses of the nucleon and the pion, respectively,
while $T_u (x,0) \equiv T(2m^2+2M^2-x, 0)$ via the isospin crossing symmetry \cite{Neville67}:
\begin{equation}
\mbox{Im} T_u^{3/2} = \frac{1}{3}\mbox{Im} T^{3/2}+\frac{2}{3}\mbox{Im} T^{1/2}~, \hspace*{0.5cm}
\mbox{Im} T_u^{1/2} = \frac{4}{3}\mbox{Im} T^{3/2}-\frac{1}{3}\mbox{Im} T^{1/2}~.
\end{equation}
To get positive definite expressions for $u$-channel amplitudes,
linear combinations of above isospin amplitudes are required.
In general, a combination $\alpha T^{1/2}+ \beta T^{3/2}$ with $0 \leq \alpha \leq 2\beta$
would lead to positivity constraints.
However, it is not difficult to show that,
it is only necessary to investigate two cases: $T^{3/2}$ and $(2T^{1/2}+T^{3/2})/3$,
which correspond to forward scattering amplitudes of $\pi^+ p \to \pi^+ p$ and $\pi^- p \to \pi^- p$, respectively.

Inside the Mandelstam triangle, $t=0$ implies that $(m-M)^2 \leq s~,u \leq (m+M)^2$.
Therefore to keep integrals in Eq. (\ref{dispersion}) positive, $n$ must be even. When $n=0$, the integrals
may be divergent and a subtraction for the scattering amplitude is needed. Therefore in our discussion,
we will choose $n=2$. Usually, higher order derivatives magnify contributions from higher dimension operators.

For convenience, one may define
\[ {\widetilde T}(\pi^+ p)= T^{3/2}-P^{3/2}~, \hspace*{0.5cm}
 {\widetilde T}(\pi^- p)=\frac{2T^{1/2}+T^{3/2}}{3}-\frac{2P^{1/2}+P^{3/2}}{3}~. \]
It is then easy to show that,
\begin{eqnarray}\label{pip}
\frac{\partial^2 {\widetilde T}(\pi^+ p)}{\partial s^2} &=&
\frac{2}{\pi}\int \limits_{(m+M)^2}^{\infty} dx \left [
\frac{\mbox{Im} T^{3/2}(x,0)}{(x-s)^3} +
\frac{2\mbox{Im} T^{1/2}(x,0)+\mbox{Im} T^{3/2}(x,0)}{3(x-u)^3}
\right ]  \\
\frac{\partial^2 {\widetilde T}(\pi^- p)}{\partial s^2} &=&
\frac{2}{\pi}\int \limits_{(m+M)^2}^{\infty} dx \left [
\frac{2\mbox{Im} T^{1/2}(x,0)+\mbox{Im} T^{3/2}(x,0)}{3(x-s)^3} +
\frac{\mbox{Im} T^{3/2}(x,0)}{(x-u)^3}
\right ] \nonumber
\end{eqnarray}
Observe that the right-hand sides of these equations are symmetric under the exchange of $s \leftrightarrow u$,
which reflects simply the crossing symmetry between $\pi^+ p$ and $\pi^- p$ scattering.
Therefore these two equations are actually not independent and one may choose freely either of them.
For definiteness, we will take $\pi^+ p$ scattering into consideration.

In Eq.~(\ref{pip}), since $\mbox{Im} T^I(x,0)$ is proportional to
the total cross section of $\pi^+ p$ scattering, the right-hand side
is positive definite, while the left-hand side can be estimated in
chiral perturbation theory ($\chi \mbox{PT}$). Usually, scattering
amplitudes in $\chi \mbox{PT}$ are expanded in powers of $\omega$,
where $\omega$ is the pion energy in the center-of-mass frame,
\[ \omega=\frac{s-m^2+M^2}{2\sqrt{s}}~. \]
Taking this into account, positivity condition from Eq.~(\ref{pip}) implies the following inequality
\begin{equation}
\left ( \frac{d \omega}{d s} \right )^2 \frac{d^2 {\widetilde T}(\pi^+ p)}{d \omega^2} +
\frac{d^2 \omega}{d s^2} \frac{d {\widetilde T}(\pi^+ p)}{d \omega} >0~.
\end{equation}

The chiral Lagrangian up to fourth order has been constructed in \cite{Fettes:2000gb},
while the corresponding scattering amplitude $T(\pi^+ p)$ has been calculated to full one loop order in heavy
baryon chiral perturbation theory (HBCPT)\cite{Fettes:2000xg}.
Analytic expressions in \cite{Fettes:2000xg} are valid in the physical region, {\it i.e.} $\omega > M$.
Inside the Mandelstam triangle, we have $-M < \omega < M$.
Therefore one should take analytic continuation of one-loop functions, for example via
\cite{Buettiker:1999ap,Bernard:1996gq},
\begin{equation}
\sqrt{\omega^2-M^2}=-i \sqrt{M^2-\omega^2}~, \hspace*{0.5cm}
\arcsin{\frac{\omega}{M}}=\frac{\pi}{2}+i\ln{\left (\frac{\omega}{M}+\sqrt{\frac{\omega^2}{M^2}-1} \right )}~.
\end{equation}
One subtle point is that $\sqrt{M^2-\omega^2}$ may appear in the
denominator of the derivative of the loop functions. Thus, to make
the series expansion well behaved, $\sqrt{1-\omega^2/M^2}$ should
not be too small. On the other hand, power expansions of scattering
amplitudes in HBCPT could break down in the limit $\omega/M \to 0$.
This is due to the presence of terms such as $M^n/\omega^m$ in
expressions of ${\widetilde T}(\pi^+ p)$, which partly reflect the
incorrect analytic behavior of chiral amplitudes inside the
Mandelstam triangle after expansion. Here the pole structures of
scattering amplitudes are changed in the heavy-baryon expansion of
propagators, hence around physical poles $s=m_N^2$ and $u=m_N^2$,
Taylor expansion of chiral amplitudes breaks down. In principle,
scattering amplitudes inside the Mandelstam triangle should be
evaluated in a relativistic formulation, such as infrared
regularization \cite{Becher:2001hv}, instead of HBCPT \footnote{We
thank the referee for pointing out this to us.}.

However, HBCPT has been widely used in phenomenological analysis of
experimental data and many of the LECs have been estimated in this
framework. It is thus desirable to stick with it in our analysis.
More so, HBCPT can actually be used if great caution has been taken
to choose an appropriate energy scale $\omega$, as we shall see
immediately. We notice that, in relatively reliable estimates on
scattering amplitudes in HBCPT, $\omega^2/M^2$ should be
significantly larger than zero to keep distance from the poles
(which is essential to the convergence of the heavy baryon
expansion), and not too close to one at the same time. To be on the
safe side, we will take $\omega = \pm M/\sqrt{2}$ in the our
numerical analysis. As a check, we have compared numerically
tree-level contributions with or without heavy baryon expansion and
found the difference to be around one percent at $\omega = \pm
M/\sqrt{2}$, which gives us some confidence about the applicability
of HBCPT with such choice of parameters. Loop level deviations are
hard to gauge, but tree-level comparisons would suggest them to be
small.

With the pion decay constant $F=92.4$ MeV, the nucleon
mass $m=938$ MeV and the pion mass $M=140$ MeV, it is
straightforward to obtain the following inequalities (LECs of small
numerical coefficients have been dropped with the understanding that
they should be of ``natural size")
\begin{eqnarray}
&&-0.6+c_1+1.2c_2-0.2 c_1 c_2-0.1c_3-0.7{\bar d}_3+0.2 ({\bar
e}_{15}+{\bar e}_{20}+{\bar e}_{35})+
\nonumber \\
&&0.6 {\bar e}_{16}+0.9{\bar d}_{18} g +(-13.4 +0.2(c_3-c_4))g^2+0.4
g^4 \gtrsim 0
\hspace*{0.5cm} \left (\omega = \frac{M}{\sqrt{2}} \right )\\
&&-0.2-c_1+1.5c_2-0.2 c_1 c_2+0.1c_3+0.9{\bar d}_3+0.2 ({\bar
e}_{15}+{\bar e}_{20}+{\bar e}_{35})+
\nonumber \\
&&0.6 {\bar e}_{16}-0.2{\bar d}_{18} g +(3.5 -0.2(c_3-c_4))g^2-0.5
g^4 \gtrsim 0 \hspace*{0.5cm} \left (\omega = -\frac{M}{\sqrt{2}}
\right )
\end{eqnarray}
Obviously, if all the LECs are taken as unknown parameters, these inequalities provide rather weak constraints.
However, by fitting to available S- and P-wave phase shifts, some of these LECs can be determined with reasonable accuracy
(see for instance Refs. \cite{Fettes:1998ud,Fettes:2000xg} and references therein).
Fitting results for the extracted values of $c_i$ and $\bar{d}_j$ are relatively stable,
therefore one might take the LECs $c_i$ and $\bar{d}_j$ as input parameters.
Then, these inequalities could potentially yield strong constraints on the $\bar{e}_k$s.

For illustration, we choose the central values of the ``Fit 2" results of \cite{Fettes:1998ud} as our input parameters,
\begin{equation}
 c_1=-1.42~,~~~c_2=3.13~, ~~~c_3=-5.85~, ~~~c_4=3.50~, ~~~{\bar d}_3=-2.75~, ~~~{\bar d}_{18}=-0.78~.
\end{equation}
With $g=1.26$, one obtain
\begin{eqnarray}
0.2 ({\bar e}_{15}+{\bar e}_{20}+{\bar e}_{35})+0.6 {\bar e}_{16}
\gtrsim 18.7 ; && \left (\omega = \frac{M}{\sqrt{2}} \right )
\label{inequality}\\
0.2 ({\bar e}_{15}+{\bar e}_{20}+{\bar e}_{35})+0.6 {\bar e}_{16}
\gtrsim -11.3; && \left (\omega = -\frac{M}{\sqrt{2}} \right )
\end{eqnarray}
Although one knows little about the ${\bar e}_k$s, the above inequalities,
especially Eq.~(\ref{inequality}), seems to be nontrivial,
which implies that ${\bar e}_{15,16,20,35}$ should be the order of ten,
fairly large numbers compared to their expected ``natural size".

In summary, we have investigated dispersion relations for $\pi N$ scattering at $t=0$.
Using the optical theorem and the isospin crossing symmetry,
dispersion relations can be re-interpreted as positivity constraints on pion-nucleon scattering amplitudes and
their even-order derivatives inside the Mandelstam triangle.
HBCPT was then used (with great caution) to estimate scattering amplitudes in this unphysical region,
which in turn yield constraints on the LECs.
If some of the LECs, for example $c_i$ and $\bar{d}_j$,
can be obtained from experimental data \cite{Fettes:1998ud,Fettes:2000xg},
fairly non-trivial constraints can be obtained on other LECs $\bar{e}_{15,16,20,35}$.
On this respect, one big question is the reliability of the fitting values of $c_i$ and $\bar{d}_j$.
A related problem is the effect of operators with chiral dimension-5 or even higher,
which is also closely related to the convergence of scattering amplitudes in HBCPT.
Compared with the effect of higher dimension operators, isospin violating effects may also warrant consideration.
But none of these considerations might be easily carried out.
Any improvement in these directions will be welcome, which will lay a more solid ground for our numerical analysis.

{\bf Acknowledgements:} This work is supported in part by the National Science Foundation of China
 under grant No. 10425525 and No. 10645001.

\end{document}